\documentclass[aps,prl,twocolumn,superscriptaddress,showpacs]{revtex4}
\usepackage{epsfig}
\usepackage{graphicx}
\usepackage{dcolumn}
\usepackage{amsmath}
\usepackage{xspace}
\newcommand{\ppbar}{p{\bar p}}
\newcommand{\invpb}{\rm pb^{-1}} 
\newcommand{\bfinvpb}{\bf pb^{-1}} 
\newcommand{\roots}{{\sqrt s}}
 
\newcommand{\Et}{\rm E_T}
\newcommand{\Pt}{\rm p_T}

\newcommand{\eeggmet}{ee\gamma\gamma\met}
\newcommand{\lgal}{\ell\gamma}
\newcommand{\lgX}{\ell\gamma\plus X}
\newcommand{\ggX}{\gamma\gamma+X}

\newcommand{\gt}{>}

\newcommand{\met}{{\rm\not\!\!E}_{T}}
\newcommand{\metvec}{{\not\!\! \vec{E}_T}}
\newcommand{\lepvec}{{\vec{E}_T^\ell}}
\newcommand{\phovec}{{\vec{E}_T^\gamma}}

\newcommand{\lplm}{\ell^+\ell^-}
\newcommand{\lgmet}{\ell\gamma\met}

\newcommand{\eg}{e\gamma}
\newcommand{\mug}{\mu\gamma}

\newcommand{\llg}{\ell\ell\gamma}
\newcommand{\llgg}{\ell\ell\gamma\gamma}
\newcommand{\eeg}{ee\gamma}
\newcommand{\mumug}{\mu\mu\gamma}
\newcommand{\Zee}{Z^0 \rightarrow e^+e^-}

\newcommand{\Wg}{W \gamma}
\newcommand{\Zg}{Z^0 \gamma}
\newcommand{\Wgg}{W \gamma\gamma}
\newcommand{\Zgg}{Z^0 \gamma\gamma}

\newcommand{\Zgstar}{Z^0\kern -0.25em/\kern -0.15em\gamma^*}
\newcommand{\epem}{e^+e^-}
\newcommand{\lumi}{305}
\newcommand{\dlumi}{18}
\newcommand{\goes}{\kern -0.18em\rightarrow\kern -0.18em}
\newcommand{\plus}{\kern -0.18em +\kern -0.18em}
\newcommand{\degs}{\mbox{$^{\circ}$}}
\newcommand{\ntotinclusive}{574}
\newcommand{\nelinclusive}{508}
\newcommand{\nmuinclusive}{66}
\newcommand{\nelbtb}{397}
\newcommand{\ntotlgammamet}{42}
\newcommand{\nelgammamet}{25}
\newcommand{\nmugammamet}{17}
\newcommand{\ntotllg}{31}
\newcommand{\nelllg}{19}
\newcommand{\nmullg}{12}
\newcommand{\MeV}{\ensuremath{\mathrm{\ Me\kern -0.1em V}}\xspace}
\newcommand{\MeVc}{\ensuremath{\mathrm{\ Me\kern -0.1em V\kern -0.1em 
/\mathit{c}}}\xspace}
\newcommand{\MeVcsq}{\ensuremath{\mathrm{\ Me\kern -0.1em V\kern -0.1em 
/\mathit{c}^2}}\xspace}
\newcommand{\GeV}{\ensuremath{\mathrm{Ge\kern -0.1em V}}\xspace}
\newcommand{\GeVc}{\ensuremath{\mathrm{\ Ge\kern -0.1em V\kern -0.1em 
/\mathit{c}}}\xspace}
\newcommand{\GeVcsq}{\ensuremath{\mathrm{\ Ge\kern -0.1em V\kern -0.1em 
/\mathit{c}^2}}\xspace}
\newcommand{\TeV}{\ensuremath{\mathrm{Te\kern -0.1em V}}\xspace}
\newcommand{\bfTeV}{\ensuremath{\bf{Te\kern -0.1em V}}\xspace}
\newcommand{\nsGeV}{\ensuremath{\mathrm{Ge\kern -0.1em V}}\xspace}
\newcommand{\nsGeVc}{\ensuremath{\mathrm{Ge\kern -0.1em V\kern -0.1em 
/\mathit{c}}}\xspace}
\newcommand{\Etgamma}{\ensuremath{\mathrm{E_T^{\gamma}}}}
\newcommand{\Etlepton}{\ensuremath{\mathrm{E_T^{\ell}}}}

\begin{document}
\title{Search for New Physics in Lepton + Photon + X
 Events with $\bf\lumi$ $\bfinvpb$ of $\bf\ppbar$ Collisions at $\bf\roots$= 1.96 $\bfTeV$}
\affiliation{Institute of Physics, Academia Sinica, Taipei, Taiwan 11529, Republic of China} 
\affiliation{Argonne National Laboratory, Argonne, Illinois 60439} 
\affiliation{Institut de Fisica d'Altes Energies, Universitat Autonoma de Barcelona, E-08193, Bellaterra (Barcelona), Spain} 
\affiliation{Baylor University, Waco, Texas  76798} 
\affiliation{Istituto Nazionale di Fisica Nucleare, University of Bologna, I-40127 Bologna, Italy} 
\affiliation{Brandeis University, Waltham, Massachusetts 02254} 
\affiliation{University of California, Davis, Davis, California  95616} 
\affiliation{University of California, Los Angeles, Los Angeles, California  90024} 
\affiliation{University of California, San Diego, La Jolla, California  92093} 
\affiliation{University of California, Santa Barbara, Santa Barbara, California 93106} 
\affiliation{Instituto de Fisica de Cantabria, CSIC-University of Cantabria, 39005 Santander, Spain} 
\affiliation{Carnegie Mellon University, Pittsburgh, PA  15213} 
\affiliation{Enrico Fermi Institute, University of Chicago, Chicago, Illinois 60637} 
\affiliation{Joint Institute for Nuclear Research, RU-141980 Dubna, Russia} 
\affiliation{Duke University, Durham, North Carolina  27708} 
\affiliation{Fermi National Accelerator Laboratory, Batavia, Illinois 60510} 
\affiliation{University of Florida, Gainesville, Florida  32611} 
\affiliation{Laboratori Nazionali di Frascati, Istituto Nazionale di Fisica Nucleare, I-00044 Frascati, Italy} 
\affiliation{University of Geneva, CH-1211 Geneva 4, Switzerland} 
\affiliation{Glasgow University, Glasgow G12 8QQ, United Kingdom} 
\affiliation{Harvard University, Cambridge, Massachusetts 02138} 
\affiliation{Division of High Energy Physics, Department of Physics, University of Helsinki and Helsinki Institute of Physics, FIN-00014, Helsinki, Finland} 
\affiliation{University of Illinois, Urbana, Illinois 61801} 
\affiliation{The Johns Hopkins University, Baltimore, Maryland 21218} 
\affiliation{Institut f\"{u}r Experimentelle Kernphysik, Universit\"{a}t Karlsruhe, 76128 Karlsruhe, Germany} 
\affiliation{High Energy Accelerator Research Organization (KEK), Tsukuba, Ibaraki 305, Japan} 
\affiliation{Center for High Energy Physics: Kyungpook National University, Taegu 702-701, Korea; Seoul National University, Seoul 151-742, Korea; and SungKyunKwan University, Suwon 440-746, Korea} 
\affiliation{Ernest Orlando Lawrence Berkeley National Laboratory, Berkeley, California 94720} 
\affiliation{University of Liverpool, Liverpool L69 7ZE, United Kingdom} 
\affiliation{University College London, London WC1E 6BT, United Kingdom} 
\affiliation{Centro de Investigaciones Energeticas Medioambientales y Tecnologicas, E-28040 Madrid, Spain} 
\affiliation{Massachusetts Institute of Technology, Cambridge, Massachusetts  02139} 
\affiliation{Institute of Particle Physics: McGill University, Montr\'{e}al, Canada H3A~2T8; and University of Toronto, Toronto, Canada M5S~1A7} 
\affiliation{University of Michigan, Ann Arbor, Michigan 48109} 
\affiliation{Michigan State University, East Lansing, Michigan  48824} 
\affiliation{Institution for Theoretical and Experimental Physics, ITEP, Moscow 117259, Russia} 
\affiliation{University of New Mexico, Albuquerque, New Mexico 87131} 
\affiliation{Northwestern University, Evanston, Illinois  60208} 
\affiliation{The Ohio State University, Columbus, Ohio  43210} 
\affiliation{Okayama University, Okayama 700-8530, Japan} 
\affiliation{Osaka City University, Osaka 588, Japan} 
\affiliation{University of Oxford, Oxford OX1 3RH, United Kingdom} 
\affiliation{University of Padova, Istituto Nazionale di Fisica Nucleare, Sezione di Padova-Trento, I-35131 Padova, Italy} 
\affiliation{LPNHE, Universite Pierre et Marie Curie/IN2P3-CNRS, UMR7585, Paris, F-75252 France} 
\affiliation{University of Pennsylvania, Philadelphia, Pennsylvania 19104} 
\affiliation{Istituto Nazionale di Fisica Nucleare Pisa, Universities of Pisa, Siena and Scuola Normale Superiore, I-56127 Pisa, Italy} 
\affiliation{University of Pittsburgh, Pittsburgh, Pennsylvania 15260} 
\affiliation{Purdue University, West Lafayette, Indiana 47907} 
\affiliation{University of Rochester, Rochester, New York 14627} 
\affiliation{The Rockefeller University, New York, New York 10021} 
\affiliation{Istituto Nazionale di Fisica Nucleare, Sezione di Roma 1, University of Rome ``La Sapienza," I-00185 Roma, Italy} 
\affiliation{Rutgers University, Piscataway, New Jersey 08855} 
\affiliation{Texas A\&M University, College Station, Texas 77843} 
\affiliation{Istituto Nazionale di Fisica Nucleare, University of Trieste/\ Udine, Italy} 
\affiliation{University of Tsukuba, Tsukuba, Ibaraki 305, Japan} 
\affiliation{Tufts University, Medford, Massachusetts 02155} 
\affiliation{Waseda University, Tokyo 169, Japan} 
\affiliation{Wayne State University, Detroit, Michigan  48201} 
\affiliation{University of Wisconsin, Madison, Wisconsin 53706} 
\affiliation{Yale University, New Haven, Connecticut 06520} 
\author{A.~Abulencia}
\affiliation{University of Illinois, Urbana, Illinois 61801}
\author{D.~Acosta}
\affiliation{University of Florida, Gainesville, Florida  32611}
\author{J.~Adelman}
\affiliation{Enrico Fermi Institute, University of Chicago, Chicago, Illinois 60637}
\author{T.~Affolder}
\affiliation{University of California, Santa Barbara, Santa Barbara, California 93106}
\author{T.~Akimoto}
\affiliation{University of Tsukuba, Tsukuba, Ibaraki 305, Japan}
\author{M.G.~Albrow}
\affiliation{Fermi National Accelerator Laboratory, Batavia, Illinois 60510}
\author{D.~Ambrose}
\affiliation{Fermi National Accelerator Laboratory, Batavia, Illinois 60510}
\author{S.~Amerio}
\affiliation{University of Padova, Istituto Nazionale di Fisica Nucleare, Sezione di Padova-Trento, I-35131 Padova, Italy}
\author{D.~Amidei}
\affiliation{University of Michigan, Ann Arbor, Michigan 48109}
\author{A.~Anastassov}
\affiliation{Rutgers University, Piscataway, New Jersey 08855}
% ...
\collaboration{CDF Collaboration}
\noaffiliation

\preprint{FERMILAB-PUB-05/yyy-E}

\date{\today}

\begin{abstract}
We present results of a search for anomalous production of events
containing a charged lepton ($\ell$, either $e$ or $\mu$) and a photon
($\gamma$), both with high transverse momentum, accompanied by
additional signatures, X, including missing transverse energy ($\met$)
and additional leptons and photons. We use the same selection criteria
as in a previous CDF search, but with a substantially larger data set,
$\lumi$ $\invpb$, a $\ppbar$ collision energy of 1.96 $\TeV$, and the
CDF II detector. We find 42 $\lgmet$ events versus an expectation of
37.3 $\pm$ 5.4 events. We observe $\ntotllg$ $\llg + X$ events versus
an expectation of 23.0 $\pm$ 2.7 events. We find no events similar to
the Run I $\eeggmet$ event.
\end{abstract}

\pacs{13.85.Rm, 12.60.Jv, 13.85.Qk, 14.80.Ly}

\maketitle

In 1995 the CDF experiment, studying $\ppbar$ collisions in 86
$\invpb$ of data at a center-of-mass energy of 1.8 $\TeV$ at the
Fermilab Tevatron, observed~\cite{Toback_all}
an event consistent with the production of two energetic photons, two
energetic electrons, and large missing transverse energy
$\met$~\cite{EtPt}. This signature is predicted to be very rare in the
standard model (SM) of particle physics~\cite{SM}, with the dominant
contribution being from the production of four gauge bosons: two W
bosons and two photons. The event raised theoretical interest,
however, as the $\llgg$ signature is expected in some models of
physics ``beyond the standard model'' such as gauge-mediated models of
supersymmetry~\cite{susy} or the production of a pair of excited
electrons~\cite{excited_electron}. The detection of this event
led to the development of ``signature-based'' inclusive searches to
cast a wider net for new phenomena: in this case one search for two photons + X ($\ggX$)
~\cite{Toback_all}, and a second for one
lepton + one photon + X ($\lgX$)~\cite{Jeff_PRD,Jeff_PRL,Jeff_thesis},
where X can be $e$, $\mu$, $\gamma$, or
$\met$, plus any number of jets.

Neither Run I search revealed convincing evidence for new
physics. However, in the $\lgX$ search, the results were
consistent with SM expectations, with ``the
possible exception of photon-lepton events with large $\met$, for
which the observed total was 16 events and the SM expectation was 7.6
$\pm$ 0.7 events, corresponding in likelihood to a 2.7 sigma
effect.''~\cite{Jeff_PRL}. The Run I paper concluded: ``However, an
excess of events with 0.7\% likelihood (equivalent to 2.7 standard
deviations for a Gaussian distribution) in one subsample among the
five studied is an interesting result, but it is not a compelling
observation of new physics. We look forward to more data in the
upcoming run of the Fermilab Tevatron.''~\cite{Jeff_PRL}. In this
Letter we report the results of repeating the $\lgX$ search with the
same kinematic selection criteria in a substantially larger data set,
$\lumi\pm\dlumi$ $\invpb$, a higher $\ppbar$ collision energy, 1.96 $\TeV$, and
the CDF II detector~\cite{CDFII}.

The CDF II detector is a cylindrically symmetric spectrometer designed
to study $\ppbar$ collisions at the Fermilab Tevatron based on the
same solenoidal magnet and central calorimeters as the CDF I
detector~\cite{CDFI} from which it was upgraded. Because the analysis
described here is intended to repeat the Run I search as closely as
possible, we note especially the differences from the CDF I detector
relevant to the detection of leptons, photons, and $\met$. The
tracking systems used to measure the momenta of charged particles have
been replaced with a central outer tracker (COT) with smaller drift
cells~\cite{COT}, and an enhanced system of silicon strip
detectors~\cite{SVX}. The calorimeters in the regions~\cite{CDF_coo}
with pseudorapidity $|\eta| \gt 1$ have been replaced with a more
compact scintillator-based design, retaining the projective
geometry~\cite{cal_upgrade}. 
The coverage in $\varphi$ of the CMP and CMX muon
systems~\cite{muon_systems} has been extended; the CMU system is
unchanged~\cite{CDFII}.

A 3-level trigger~\cite{CDFII} system selects events 
with a high transverse momentum ($\Pt$)~\cite{EtPt} lepton ($\Pt >
18~\GeV$) or
photon ($\Et > 25~\GeV$) in the central region, $|\eta|
\lesssim 1.0$. The trigger system selects photon and electron
candidates from clusters of energy in the central electromagnetic
calorimeter. Electrons are distinguished from photons by
requiring a COT track pointing at the cluster. The
muon trigger requires a COT track that extrapolates to a 
track segment (``stub'') in the muon chambers.

Inclusive $\lgal$ events are selected by requiring a central $\gamma$
candidate with $\Etgamma>25$ $\GeV$ and a central $e$
or $\mu$ with $\Etlepton>25$ $\GeV$ originating 
less than 60 cm along the beam-line from the detector center and
passing the ``tight'' criteria listed below.

The identification of leptons and photons is essentially the same as
in the Run I search~\cite{Jeff_PRD}.
A muon candidate passing the ``tight'' cuts must have: a) a
well-measured track in the COT; b) energy deposited in the calorimeter
consistent with expectations; c) a muon ``stub'' in both the CMU and
CMP, or in the CMX, consistent with the extrapolated COT track; and d)
COT timing consistent with a track from a $\ppbar$ collision. An
electron candidate passing the ``tight'' selection must have: a) a
high-quality track with $\Pt>0.5~\Et$, unless $\Et > 100$ $\GeV$, in
which case the $\Pt$ threshold is set to 25 $\GeV$; b) a good
transverse shower profile that matches the extrapolated track
position; c) a lateral sharing of energy in the two calorimeter towers
containing the electron shower consistent with that expected; and d)
minimal leakage into the hadron calorimeter~\cite{hadoem}.

Photon candidates are required to have no track with $\Pt>1$ $\GeV$,
and at most one track with $\Pt<1$ $\GeV$, pointing at the calorimeter
cluster; good profiles in both transverse dimensions at shower
maximum; and minimal leakage into the hadron
calorimeter~\cite{hadoem}.

To reduce background from photons or leptons from the decays of
hadrons produced in jets, both the photon and the lepton in each event
are required to be ``isolated''. The $\Et$ deposited in the
calorimeter towers in a cone in $\eta-\varphi$ space~\cite{CDF_coo} of
radius $R=0.4$ around the photon or lepton position is summed, and the
$\Et$ due to the photon or lepton is subtracted. The remaining $\Et$
is required to be less than
$2.0~\GeV+0.02\times(\Et-20~\GeV)$ for a photon, or less than 10\% of
the $\Et$ for electrons or $\Pt$ for muons. In addition, for photons
the sum of the $\Pt$ of all tracks in the cone must be less than
$2.0~\GeV+0.005\times\Et$.

Missing transverse energy $\met$ is calculated from the calorimeter
tower energies in the region $|\eta| < 3.6$. Corrections are then made
to the $\met$ for non-uniform calorimeter response~\cite{jet_corr} for
jets with uncorrected $\Et > 15$ $\GeV$ and $\eta < 2.0$, and for
muons with $\Pt > 20$ $\GeV$.

A total of $\ntotinclusive$ events, $\nelinclusive$ inclusive $\eg$
and $\nmuinclusive$ inclusive $\mug$ candidates, pass the $\lgal$
selection criteria. Of the $\nelinclusive$ inclusive $\eg$ events,
$\nelbtb$ have the electron and photon within 30$\degs$ of
back-to-back in $\varphi$, $\met< 25$ $\GeV$, and no additional
leptons or photons. These are dominated by $\Zee$ decays in which one
of the electrons radiates a high-$\Et$ photon while traversing the
material inside the COT active volume, leading to the observation of
an electron and a photon approximately back-to-back in $\varphi$, with
an $\eg$ invariant mass close to the $Z^0$ mass.

\begin{figure}[!tb]
\begin{center}
\hspace*{-0.1in}
\includegraphics*[angle =90,width=0.50\textwidth]{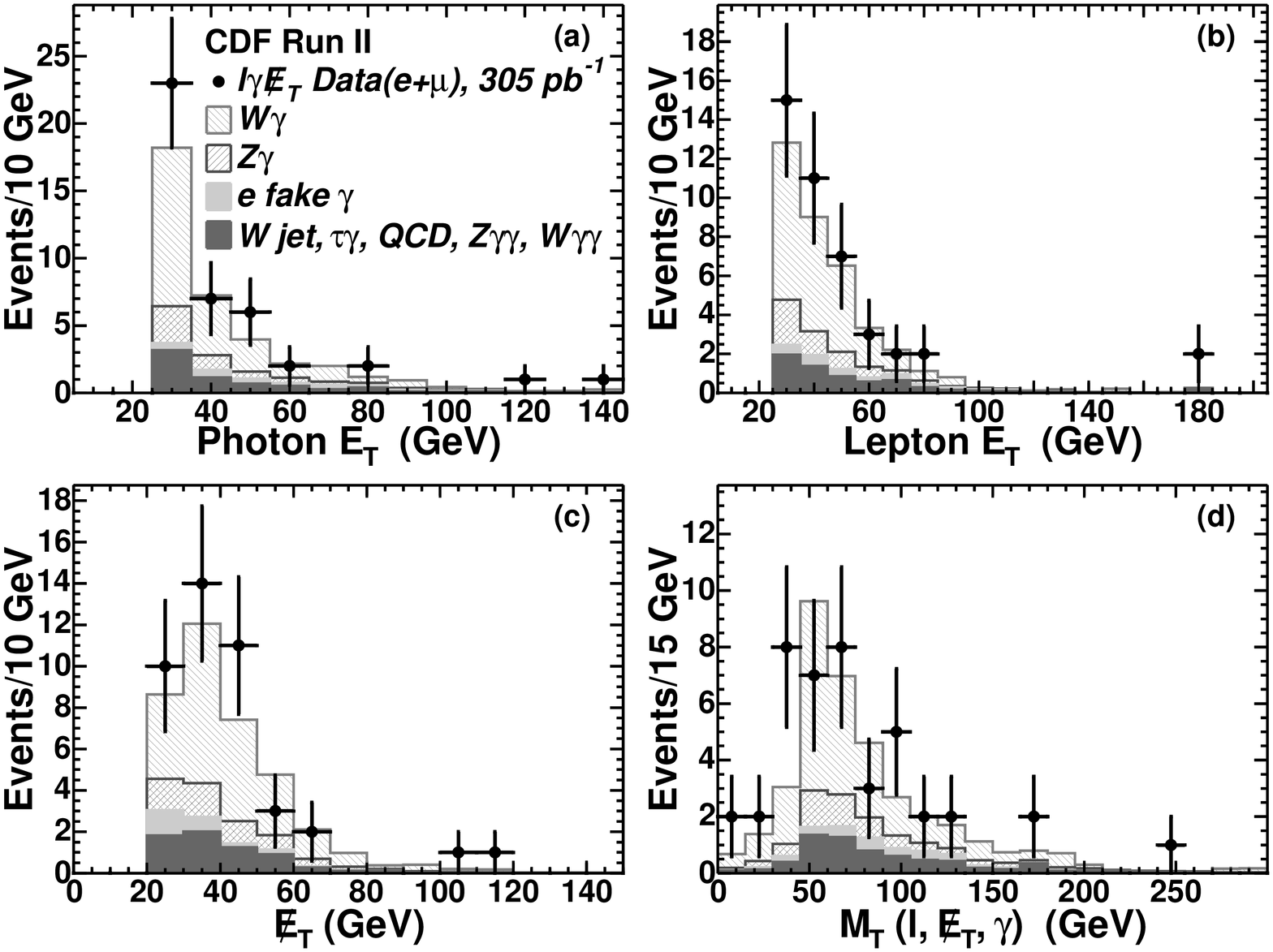}
\end{center}
\caption{ The distributions for events in the $\lgmet$ sample (points)
  in a) the $\Et$ of the photon; b) the $\Et$ of the lepton; c) the
  missing transverse energy, $\met$; and d) the transverse mass of the
  $\lgmet$ system.  The histograms
show the expected SM
  contributions, including estimated backgrounds from misidentified
  photons and leptons.}
\label{figure1}
\end{figure}

We use $W^\pm$ and $Z^0$ production as control samples to ensure that
the efficiencies for high-$\Pt$ electrons and muons, as well as for
$\met$, are well understood. The photon control sample is constructed
from events in which one of the electrons radiates a high-$\Et$
$\gamma$ such that the $\eg$ invariant mass
is within 10 $\GeV$ of the $Z^0$ mass.

The first search we perform is in the $\lgmet+X$ subsample, defined by
requiring that an event contain $\met> 25~\GeV$ in addition to the
$\gamma$ and ``tight'' lepton. Of the $\ntotinclusive$ $\lgal$ events,
$\nelgammamet$ $e\gamma\met$ events and $\nmugammamet$ $\mu\gamma\met$
events pass the $\met$ requirement. Figure~\ref{figure1} shows the
observed distributions 
in a) the $\Et$ of the photon; b) the $\Et$ of the lepton; c) $\met$; 
and d) the transverse mass of the
$\lgmet$ system, where 
 $\rm{M_T} = [{(\rm E_T^\ell+E_T^\gamma + \met)^2}$ -
$(\lepvec + \phovec + \metvec)^2]^{1/2}$.

A second search, for the $\llg+X$ signature, is constructed by
requiring another $e$ or $\mu$ in addition to the ``tight'' lepton
and the $\gamma$. The additional muons are required to have $\Pt >
20~\GeV$ and to satisfy the same 
criteria as for ``tight'' muons but with fewer
hits required on the track, or, alternatively,
 a more stringent cut on track quality
but no requirement that there be a matching ``stub'' in the muon
systems. Additional central electrons are required to have $\Et >
20~\GeV$ and to satisfy the tight central electron criteria
but with a track requirement of only $\Pt>10$ $\GeV$ (rather than
0.5$\times\Et$), and no requirement on a shower maximum measurement or
lateral energy sharing between calorimeter towers. Electrons in the
end-plug calorimeters ($1.2 < |\eta| < 2.0$) are required to have
$\Et> 15~\GeV$, minimal leakage into the hadron calorimeter, a
``track'' containing at least 3 hits in the silicon tracking system,
and a shower transverse shape consistent with that expected, with a
centroid close to the extrapolated position of the
track~\cite{wenu_asymmetry_paper}.

\begin{figure}[b]
\vspace*{-0.2in}
\begin{center}
\hspace*{-0.1in}
\includegraphics*[angle=90,width=0.50\textwidth]{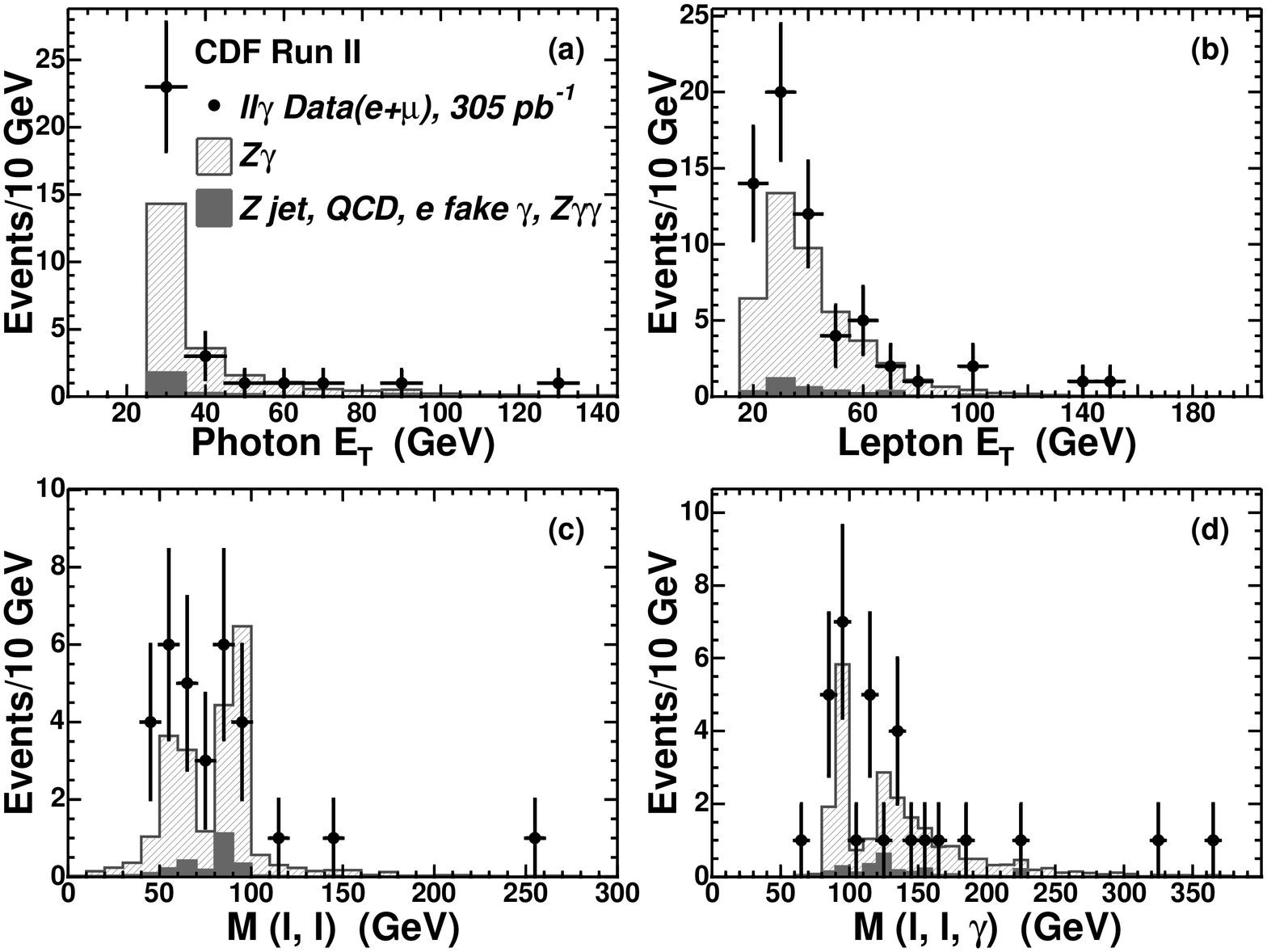}
\end{center}
\caption{ The distributions for events in the $\llg$ sample (points)
in a) the $\Et$ of the photon; b) the $\Et$ of the leptons (two
entries per event); c) the 2-body mass of the dilepton system; and d)
the 3-body mass $m_{\llg}$. The histograms show the expected SM
contributions.}
\label{figure2}
\end{figure}

The $\llg$ search criteria select $\ntotllg$ events ($\nelllg$ $\eeg$
and $\nmullg$ $\mumug$) of the $\ntotinclusive$ $\lgal$ events. No
$e\mu\gamma$ events are observed. Figure~\ref{figure2} shows the
observed distributions in a) the $\Et$ of the photon; b) the $\Et$ of
the leptons; c) the 2-body mass of the dilepton system; and d) the
3-body mass $m_{\llg}$.

\begin{figure}[!b]
\vspace*{-0.2in}
\begin{center}
\includegraphics*[width=0.18\textwidth, angle=90,clip=]{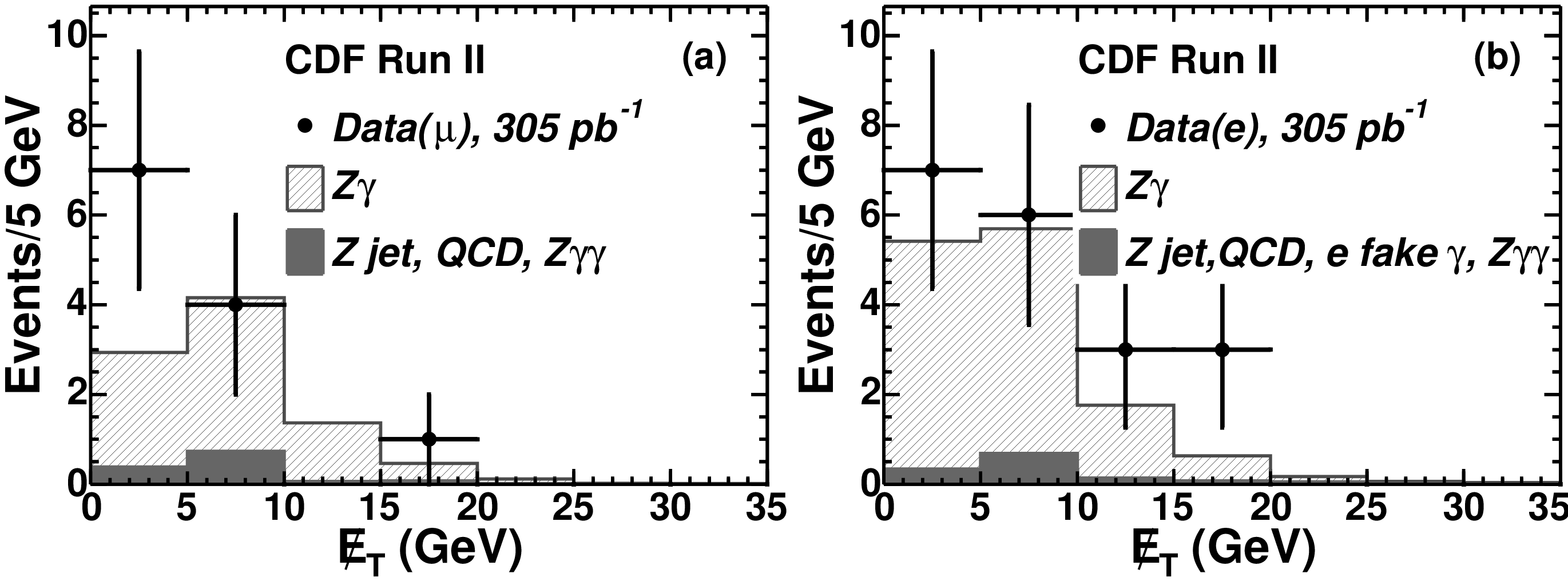}
\end{center}
\caption{The distributions in missing transverse energy $\met$
  observed in the inclusive search for a) $\mumug$ events and
  b) $\eeg$ events. The histograms show the expected SM
  contributions.}
\label{figure3}
\end{figure}

We do not expect SM events with large $\met$ in the $\llg$ sample; 
the Run I $\eeggmet$ event was of special
interest in the context of supersymmetry~\cite{susy} due to the large
value of $\met$ (55 $\pm$ 7 $\GeV$). Figure~\ref{figure3} shows the
distributions in $\met$ for the $\mumug$ and $\eeg$ subsamples of the
$\llg$ sample. No events are observed with $\met> 25$ $\GeV$.

The dominant SM source of $\lgal$ events is electroweak
$W$ and $\Zgstar$ production along with a $\gamma$ radiated from one
of the charged particles involved in the process~\cite{CDF_WZgamma}.
The number of such events is estimated
using leading-order (LO) event
generators~\cite{MadGraph,Baur,CompHep}. Initial state radiation is
simulated by the {\sc pythia} shower Monte Carlo (MC) code~\cite{Pythia}
tuned to reproduce the underlying event. The generated particles
are then passed through a full detector simulation, and these
events are then reconstructed with the same code used
for the data.

The expected contributions from $\Wg$ and $\Zgstar\plus\gamma$
production to the $\lgmet$ and $\llg$ searches are given in
Table~\ref{summary_table}. A correction for higher-order processes (K-factor) 
has been applied~\cite{Baur_NLO}. In the $\lgmet$ signature we expect
22.5 $\pm$ 2.8 events from $\Wg$ and 5.7 $\pm$ 1.0 from
$\Zgstar\plus\gamma$. In the $\llg$ signature, we expect
20.3 $\pm$ 2.4 events from $\Zgstar\plus\gamma$; the contribution from
$\Wg$ is negligible. The uncertainties on the SM contributions include
those from parton distribution functions (5\%), factorization scale
(2\%), and K-factor (3\%), a comparison of different MC generators
($\sim$ 5\%), and the luminosity (6\%).

\begin{table}[!h]
\begin{center}
\caption{A comparison of the numbers of events predicted by the
SM and the observations for the $\lgmet$ and $\llg$ searches. The SM
predictions are dominated by $\Wg$ and $\Zg$
production, respectively~\cite{MadGraph,Baur,CompHep}. Other
contributions come from $\Wgg$ and $\Zgg$,
leptonic $\tau$ decays, and misidentified leptons, photons, or
$\met$.}
\label{summary_table} 
\begin{tabular}{lccc}

\hline
\multicolumn{4}{c} {
{\bf Lepton+Photon+$\bf\met$ Events}}\\
\hline
{\bf SM Source} & {\bf   $e\gamma\met$} & {\bf   $\mu\gamma\met$} & {\bf   $(e+\mu)\gamma\met$} \\
\hline
$W^{\pm}\gamma$                         & 
13.70$\pm$1.89    &
 8.84$\pm$1.35    &
22.54$\pm$2.80    \\
$\Zgstar + \gamma$                              & 
1.16$\pm$0.40     &
4.49$\pm$0.64    &
5.65$\pm$1.03     \\
$W^{\pm}\gamma\gamma, \Zgstar \plus \gamma\gamma$   & 
0.14$\pm$0.02   &
0.18$\pm$0.02   &
0.32$\pm$0.03  \\
$W^{\pm}\gamma,\Zgstar\plus\gamma\goes\tau\gamma$     & 
0.71$\pm$0.18  &
0.26$\pm$0.08  &
0.97$\pm$0.22  \\
\hline
$W^{\pm}$+Jet faking $\gamma$                     & 
2.8$\pm$2.8       &
1.6$\pm$1.6       &
4.4$\pm$4.4       \\
$\Zgstar \goes \epem,e\goes\gamma$           & 
2.45$\pm$0.33    &
-                                       &
2.45$\pm$0.33                           \\
Jets faking $\ell+\met$                         & 
0.7$\pm$0.7   &
0.3$\pm$0.3   &
1.0$\pm$0.8   \\
\hline
{\bf Total} & 
{\bf 21.7$\pm$3.4}     & 
{\bf 15.7$\pm$2.2}     & 
{\bf 37.3$\pm$5.4}     \\

\hline
{\bf Observed} & 
{\bf   25}                          & 
{\bf   17}                          & 
{\bf   42}                          \\
\hline

\multicolumn{4}{c}{
{\bf Multi-Lepton+Photon Events}}\\ 
\hline
{\bf SM Source} & {\bf  $ee\gamma$} & {\bf  $\mu\mu\gamma$} & {\bf  $ll\gamma$}  \\
\hline
$\Zgstar + \gamma$            		          & 
12.50$\pm$1.53    &
 7.81$\pm$0.88    &
20.31$\pm$2.40    \\
$\Zgstar + \gamma\gamma$ 	     	& 
0.24$\pm$0.03  &
0.12$\pm$0.02 &
0.36$\pm$0.04  \\
\hline
$\Zgstar+$Jet faking $\gamma$   & 
0.3$\pm$0.3     &
0.2$\pm$0.2     &
0.5$\pm$0.5     \\
$\Zgstar \goes \epem,e\goes\gamma$           & 
0.23$\pm$0.09    &
-                                       &
0.23$\pm$0.09                           \\
Jets faking $\ell+\met$                         & 
0.6$\pm$0.6  &
1.0$\pm$1.0  &
1.6$\pm$1.2  \\
\hline
{\bf Total} & 
{\bf 13.9$\pm$1.7}    & 
{\bf  9.1$\pm$1.4}    & 
{\bf 23.0$\pm$2.7}   \\

\hline
{\bf Observed} & 
{\bf   19}                          & 
{\bf   12}                          &                 
{\bf   31}                          \\
\hline
\end{tabular}
\end{center}
\end{table}

High $\Pt$ photons are copiously created from hadron decays in jets
initiated by a scattered quark or gluon. In particular, mesons such as
the $\pi^0$ or $\eta$ decay to photons which may satisfy the
photon selection criteria. The numbers of
lepton-plus-misidentified-jet events expected in the $\lgmet$ and
$\llg$ samples are determined by measuring the jet $\Et$ spectrum in
$\ell\,\met+$jet and $\ell\ell+$jet samples, respectively, and then
multiplying by the probability of a jet being misidentified as a
photon, $P^{jet}_{\gamma}(\Et)$, which is measured in data samples
triggered on jets. 
The misidentification rate is
$P^{jet}_{\gamma}=(6.5~\pm~3.3)\times10^{-4}$ for $\Et^{\gamma} = 25$
$\GeV$, and $(4.0~\pm~4.0)\times10^{-4}$ for $\Et^{\gamma} = 50$
$\GeV$ ~\cite{CDF_WZgamma}.
The predicted number of events with jets misidentified as photons is
4.4 $\pm$ 4.4 for the $\lgmet$ signature and 0.5 $\pm$ 0.5 for
$\llg$.

The probability that an electron undergoes hard bremsstrahlung
and is misidentified as a photon, $P^{e}_{\gamma}$, is measured from the
photon control sample.
The number of
misidentified $\eg$ events divided by twice the number of
$ee$ events gives $P^{e}_{\gamma}$=(1.7 $\pm$ 0.1)\%.
Applying this misidentification rate to electrons in the inclusive
lepton samples, we find 2.5 $\pm$ 0.3 and 0.2 $\pm$ 0.1 events pass
the selection criteria for the $\lgmet$ and $\llg$ searches,
respectively.

We have estimated the background due to events with jets misidentified
as $\lgmet$ or $\llg$ signatures by studying the total $\Pt$ of tracks
in a cone in $\eta-\varphi$ space of radius $R=0.4$ around the lepton
track. We estimate there are 1.0 $\pm$ 0.8 and 1.6 $\pm$ 1.2 events in
the $\lgmet$ and $\llg$ signatures, respectively.

We have used both {\sc madgraph}~\cite{MadGraph} and {\sc comphep}\cite{CompHep}
to simulate the triboson channels $\Wgg$ and $Z\gamma\gamma$. The
expected contributions are small, 0.32 $\pm$ 0.03 and
0.36 $\pm$ 0.04 events in the $\lgmet$ and $\llg$ signatures,
respectively.

Muon backgrounds from hadrons either decaying in-flight or penetrating
the iron before the muon chambers, and from the decay of bottom and
charm quarks, are found to be negligible.

The predicted and observed totals for both the $\lgmet$ and $\llg$
searches are shown in Table~\ref{summary_table}. We observe
$\ntotlgammamet$ $\lgmet$ events, versus the expectation of 37.3 $\pm$
5.4 events. In the $\llg$ channel, we observe $\ntotllg$ events,
versus an expectation of 23.0 $\pm$ 2.7 events. There is no
significant excess in either signature. The predicted and observed
kinematic distributions are compared in Figure~\ref{figure1} for the
$\lgmet$ signature, and Figures~\ref{figure2} and \ref{figure3} for
the $\llg$ search.

In conclusion, we have repeated the search for inclusive lepton +
photon production with the same kinematic requirements as the Run I
search, but with a significantly larger data sample and a higher
collision energy. We find that the numbers of events in the $\lgmet$
and $\llg$ subsamples of the $\lgX$ sample agree with SM
predictions. We observe no $\llg$ events with anomalous large $\met$
or with multiple photons and so find no events like the $\eeggmet$
event of Run I.

\begin{acknowledgments}

We thank the Fermilab staff and the technical staffs of the
participating institutions for their vital contributions. Uli Baur,
Alexander Belyaev, Edward Boos, Lev Dudko, Tim Stelzer, and Steve
Mrenna were extraordinarily helpful with the SM predictions. This
work was supported by the U.S. Department of Energy and National
Science Foundation; the Italian Istituto Nazionale di Fisica Nucleare;
the Ministry of Education, Culture, Sports, Science and Technology of
Japan; the Natural Sciences and Engineering Research Council of
Canada; the National Science Council of the Republic of China; the
Swiss National Science Foundation; the A.P. Sloan Foundation; the
Bundesministerium f\"ur Bildung und Forschung, Germany; the Korean
Science and Engineering Foundation and the Korean Research Foundation;
the Particle Physics and Astronomy Research Council and the Royal
Society, UK; the Russian Foundation for Basic Research; the Comisi\'on
Interministerial de Ciencia y Tecnolog\'{\i}a, Spain; in part by the
European Community's Human Potential Programme under contract
HPRN-CT-2002-00292; and the Academy of Finland.
\end{acknowledgments}



\begin{thebibliography}{99}

\bibitem{Toback_all}
F.~Abe \textit{et al.} (CDF Collaboration), Phys. Rev. D \textbf{59},
092002 (1999); F.~Abe \textit{et al.} (CDF Collaboration),
Phys. Rev. Lett. \textbf{81}, 1791 (1998); D.~Toback, Ph.D. thesis,
University of Chicago, 1997.

%
%

\bibitem{EtPt} Transverse momentum and energy are defined as $\Pt =
p\sin\theta$ and $\Et = E\sin\theta$, respectively.  
%
Missing $\rm E_T$ ($\metvec$) is defined by $\metvec = -\sum_{i} E_T^i
\hat{n}_i$, where i is the calorimeter tower number for $|\eta| <$ 3.6
(see Ref.~\cite{CDF_coo}), and $\hat{n}_i$ is a unit vector
perpendicular to the beam axis and pointing at the i$^{th}$
tower. We correct $\metvec$ for jets and muons. We define
the magnitude $\met=|\metvec|$.
%
We use the convention that ``momentum'' refers to $pc$ and ``mass'' to $mc^2$.

\bibitem{SM} S.L.~Glashow,
Nucl. Phys. {\bf 22} 588, (1961); S. Weinberg,
Phys. Rev. Lett. {\bf 19} 1264, (1967);
A. Salam, Proc. 8th Nobel Symposium, Stockholm, (1979).
      
\bibitem{susy} 
S.~Ambrosanio, G.L.~Kane, G.D.~Kribs, S.P.~Martin, and S.~Mrenna,
Phys. Rev. D \textbf{55}, 1372 (1997); B.C.~Allanach, S.~Lola,
K.~Sridhar, Phys. Rev. Lett. \textbf{89}, 011801 (2002).

\bibitem{excited_electron} 
D. Acosta \textit{et al.} (CDF Collaboration), Phys. Rev. Lett. \textbf{94},
101802 (2005).

\bibitem{Jeff_PRD} 
D. Acosta \textit{et al.} (CDF Collaboration), Phys. Rev. D
\textbf{66}, 012004 (2002); hep-ex/0110015.

\bibitem{Jeff_PRL} 
D. Acosta \textit{et al.} (CDF Collaboration), Phys. Rev. Lett. \textbf{89},
041802 (2002); hep-ex/0202004.

\bibitem{Jeff_thesis}
J.~Berryhill, Ph.D. thesis, University of Chicago, 2000.

\bibitem{CDFII} 
D. Acosta \textit{et al.} (CDF Collaboration), Phys. Rev. D \textbf{71}, 032001
(2005).


\bibitem{CDFI} 
F. Abe \textit{et al.} (CDF Collaboration),
Nucl. Instrum. Methods A \textbf{271}, 387 (1988).

\bibitem{COT} 
A. Affolder \textit{et al.}, 
Nucl. Instrum. Methods A \textbf{526}, 249 (2004).

\bibitem{SVX} 
A. Sill \textit{et al.}, 
Nucl. Instrum. Methods A \textbf{447}, 1 (2000);
A. Affolder \textit{et al.},
Nucl. Instrum. Methods A \textbf{453}, 84 (2000); 
C.S. Hill,
Nucl. Instrum. Methods A \textbf{530}, 1 (2000).

\bibitem{CDF_coo} The CDF coordinate system of 
$r$, $\varphi$, and $z$ is cylindrical, with the $z$-axis along the
proton beam. The pseudorapidity is $\eta = -\ln(\tan(\theta/2))$.

\bibitem{cal_upgrade} S. Kuhlmann \textit{et al.}, 
Nucl. Instrum. Methods A \textbf{518}, 39, 2004.

\bibitem{muon_systems} 
The CMU system consists of 
gas proportional
chambers in the region $|\eta|<0.6$; the CMP system consists of
chambers after an additional meter of steel, also for
$|\eta|<0.6$. The CMX chambers cover 
$0.6<|\eta|<1.0$.



\bibitem{hadoem} 
The fraction of electromagnetic energy allowed to leak into the hadron
compartment $\rm E_{had}/E_{em}$ must be less than $\rm
0.055\plus0.00045\times E_{em}(\GeV)$ for central electrons, less than
0.05 for electrons in the end-plug calorimeters, less than max[0.125,
$\rm 0.055\plus0.00045\times E_{em}(\GeV)$] for photons.


\bibitem{jet_corr} 
A. Bhatti \textit{et al.}, submitted to 
Nucl. Instrum. Methods,
Oct. 2005; hep-ex/0510047.

\bibitem{wenu_asymmetry_paper} 
D.~Acosta \textit{et al.} (CDF Collaboration),
Phys. Rev. D \textbf{71}, 051104 (2005); hep-ex/0501023.

\bibitem{CDF_WZgamma} 
D. Acosta \textit{et al.} (CDF Collaboration), Phys. Rev. Lett. \textbf{94},
041803 (2005).

\bibitem{MadGraph} 
T. Stelzer and W. F. Long, Comput. Phys. Commun. \textbf{81}, 357
(1994); F. Maltoni and T. Stelzer, JHEP \textbf{302}, 27 (2003).

\bibitem{Baur} 
U. Baur, T. Han, and J. Ohnemus,
Phys. Rev. D \textbf{48}, 5140 (1993);
J. Ohnemus, Phys. Rev. D \textbf{47}, 940 (1993).

\bibitem{CompHep} 
E. Boos \textit{et al.} (The {\sc comphep} Collaboration),
Nucl. Instrum. Methods A \textbf{534}, 250, (2004); hep-ph/0403113.

\bibitem{Pythia} 
T.~Sjostrand, Comput. Phys. Commun. \textbf{82} (1994) 74;
S.~Mrenna, Comput. Phys. Commun. \textbf{101} (1997) 232.
 
\bibitem{Baur_NLO} U.~Baur, T.~Han and J.~Ohnemus, 
  Phys.\ Rev.\ D {\bf 48}, 5140 (1993);
  U.~Baur, T.~Han and J.~Ohnemus, 
  Phys.\ Rev.\ D {\bf 57}, 2823 (1998); hep-ph/9710416.
  Both the $\Wg$ and $Z\gamma$ K-factors are fixed at 1.36 for generated
  $\ell\nu$ masses below 76 $\GeV$ and for generated $\lplm$ masses below
  86 $\GeV$. Above the poles the K-factors grow with $\Et^{\gamma}$ to
  be 1.62 and 1.53 at $\Et^{\gamma}=100$ $\GeV$ for $\Wg$ and $Z\gamma$,
  respectively.

\end{thebibliography}
\end{document}